# Haptic Repurposing with GenAI: Transforming the Tangible World into Adaptive Haptic Interfaces in Mixed Reality

Haoyu Wang

haoyu.wang22@imperial.ac.uk

Imperial College London, Royal College of Art

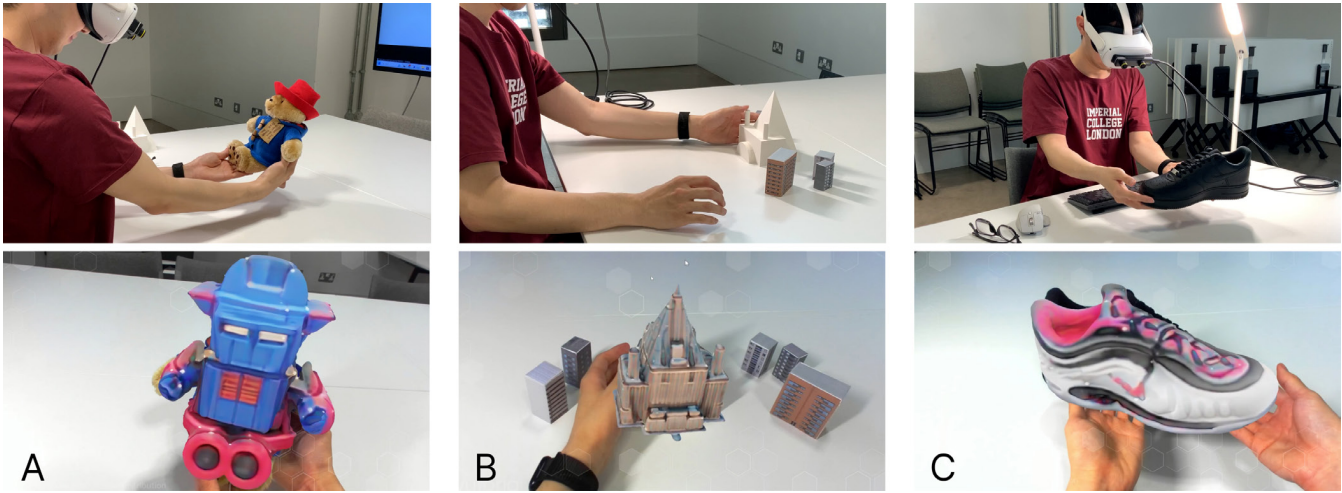

Figure 1 A:Transforming a Paddington bear toy into "a cute transformer toy"; B: Transforming a 3D printed white building model into "a Empire State Building Architecture"; C: Transforming a black shoe into "a mix of Nike Air Foamposite One with Airmax 97".

## Abstract

Mixed Reality (MR) aims to merge the digital and physical worlds to create immersive human-computer interactions. Despite notable advancements, the absence of realistic haptic feedback often breaks the immersive experience by creating a disconnect between visual and tactile perceptions. This paper introduces "Haptic Repurposing with GenAI," an innovative approach to enhance MR interactions by transforming any physical objects into adaptive haptic interfaces for AI-generated virtual assets. Utilizing state-of-the-art generative AI (GenAI) models, this system captures both 2D and 3D features of physical objects and, through user-directed prompts, generates corresponding virtual objects that maintain the physical form of the original objects. Through model-based object tracking, the system dynamically anchors virtual assets to physical props in real-time, allowing objects to visually morph into any user-specified virtual object. This paper details the system's development, presents findings from usability studies that validate its effectiveness, and explores its potential to significantly enhance interactive MR environments. The hope is this work can lay a foundation for further research into AI-driven spatial transformation in immersive and haptic technologies.

# 1 Introduction

The purpose of Mixed Reality is to create a blend of physical and digital worlds, unlocking natural and intuitive 3D human, computer, and environmental interactions[1]. With the proliferation of consumer-level head-mounted displays like the Oculus Quest series and Apple Vision Pro, the vision of blurring the line between real and virtual to create immersive experiences is becoming increasingly achievable. Despite advancements in display and audio systems, haptic feedback--a crucial element for creating truly immersive experiences--remains notably absent in many of today's MR experiences.

To address this gap, considerable efforts have been undertaken. Haptic wearables, as a hardware solution, have been extensively researched by both academia and industry [2],[3]. These devices hold promise but also present considerable challenges. Typically, these technologies often come with high costs and have not yet reached broad consumer adoption, limiting their impact within research labs. More importantly, they are limited to simulating simple tactile sensations and do not provide the complex force feedback required for more realistic interactions, such as simulating the feeling of holding an object with both hands.

Another existing approach is to use everyday physical objects as haptic props for virtual assets [4], [5]. However, given the diverse scenarios in which consumer-level headsets are used, it is impractical to find a common set of physical props that cover most probable shapes. Additionally, to avoid a sense of visual-haptic mismatch—which can negatively impact the user's immersion and engagement [6] -- the virtual assets and physical objects must share a corresponding physical form.

But what if we can dynamically transform any physical object into a variety of virtual models while preserving its original 3D structure, using existing physical objects as the haptic prop? Imagine grasping a plastic bottle that becomes a lightsaber handle in a virtual game, or petting a pillow that transforms into a crouching dragon, perfectly mirroring the pillow's physical form. In this way, users not only see transformations but also feel textures and weights, pushing the boundaries of interactivity within Mixed Reality.

To actualize this vision, we need to delve into great content within the passive haptics field, along with methodologies surrounding 3D model creating and object tracking. Haptic Repurposing with GenAI introduces a novel pipeline that leverages state-of-the-art (SOTA) generative AI models to enable Any-to-Any transformation, eliminating the dependence on specific scenarios or predefined objects. This approach greatly extends the range of applicability of passive haptics across various contexts and holds substantial potential for scalability.

## 2 Related Work

### 2.1 Passive Haptics in Mixed-Reality

The concept of using passive haptic props has been investigated by repurposing physical objects or environments to generate haptic sensations in VR and AR [5], [7], [8]. In this process, typically a mismatch between virtual and physical objects is inevitable. This could lead to a conflict in the user's mind and break the experience [9]. In the project "Substitutional Reality", Simeone et al. [6] explored the question of how large the mismatch can be before it significantly affects the believability of the experience. The result shows that some amount of mismatch is acceptable while increased mismatch negatively affects a user's believability of the visual experience.

Building on the work of Simeone et al., Annexing Reality [10] introduced a method to effectively match appropriate 3D models from a predefined set to its closest physical proxies detected in the user's environment. This enables opportunistic use of everyday common objects as tangible proxies for virtual assets. Additionally, Cathy et al.[11] use common household objects, such as chairs and sofas, as passive haptic props for preset scenarios in VR.

While these methods are impressive and novel, they are limited by their reliance on preselected 3D objects and specific scenarios. This dependency restricts their scalability and adaptability across diverse real-world settings, where user environments and available objects can vary widely.

### 2.2 Generative AI

To enable dynamic and intelligent transformation within this project, we recognized the need for an advanced method of creating 3D assets. This led us to focus on generative AI, a field that has seen significant advancements in recent years. Deep generative models have unlocked profound new realms of human creativity. By capturing and generalizing patterns within data, we have entered the epoch of all-encompassing Artificial Intelligence for General Creativity (AIGC).[12]

Among the advancements, extensive research has demonstrated notable performance improvement in the Text-to-Image generation [13], [14], [15], [16] task by leveraging pre-trained diffusion models [17], [18], [19] on a large-scale text-image dataset [20], [21]. SOTA Text-to-Image models and products such as Stable Diffusion [22], Midjourney [23] and DALL-E [24] have showcased impressive capabilities in producing images with exceptional quality and fidelity. Additionally, ControlNet [25] offers methods to incorporate spatial conditioning controls into diffusion models, enabling precise control over the image generation process.

On the other hand, advancements in 3D reconstruction from single images have been driven by the evolution of generative models [28], [29], [30], [31], [32], [33]. The development and accessibility of various open-source solutions [31], [32], [33] have made the digitization of 3D objects from 2D images feasible and swift. This advancement has substantially augmented the potential of generative AI, enabling the creation of more immersive MR experiences by seamlessly integrating AI-generated virtual objects into real-world environments.

### 2.3 Object Tracking

To ensure a more convincing transformation process, it is crucial for the virtual object to adhere closely to the real one, even during movement and rotation. This requirement underscores the importance of object tracking, a fundamental component of computer vision research. Among the various tracking methods, CAD model-based tracking is particularly notable in our context for its utilization of 3D models to enhance the accuracy and reliability of tracking performance. Among recent advances, Long et al. [34] suggest an active contour model that optimizes object tracking while achieving real-time performance on mobile devices, showcasing considerable potential. Moreover, commercial toolkits such as Vuforia [35] offer robust performance and scalability, providing essential resources for our development efforts.

The tracking solutions discussed facilitate the seamless tracking of real objects by leveraging AI-generated 3D models. This project builds upon the concepts of Annexing Reality and VR Haptics at Home. By employing the capabilities of generative AI, it expands the scope of passive haptic experiences beyond traditional limits, enabling the adaptive transformation of the real and diverse physical world into prompt-guided haptic interfaces. This innovative approach not only enhances user interaction with MR environments but also pushes the boundaries of how we perceive and interact with our surroundings.

## 3 Development Process

In this paper, we have outlined a foundational pipeline to realize the transformation (Figure 2). By detailing the methodologies employed, we aim to provide a clear framework that others in the field can easily understand, replicate, and build upon.

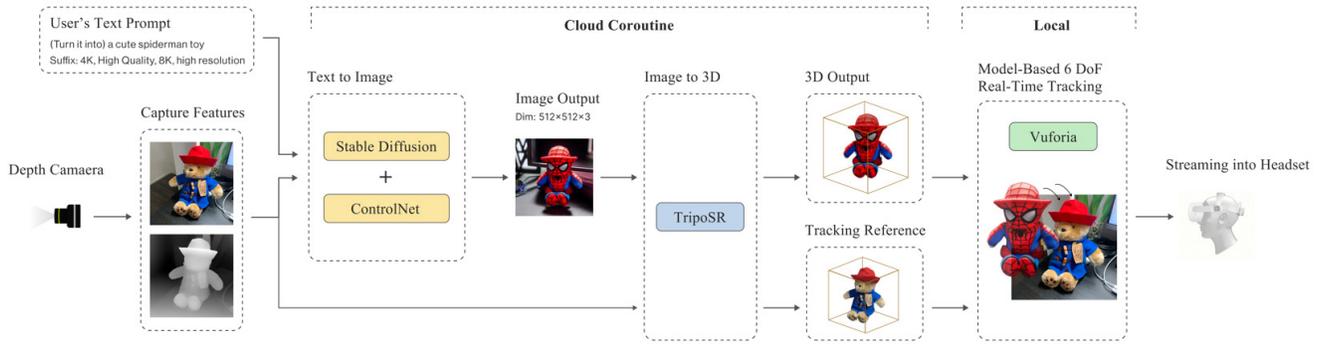

**Figure 2 Overview of the Entire Transformation Process Pipeline**

In this study, we utilized a ZED Mini depth camera [36] to capture both RGB video and depth information. This camera was strategically mounted in front of the headset as shown in Figure 1, aligned with the user's head direction to facilitate accurate data capture that corresponds with the user's natural point of view. An Oculus Quest 2 VR headset was used to stream the post-processed video, thus creating an accessible augmented reality (AR) developing environment.

To manage the computationally intensive tasks associated with generative AI, we built a cloud server using Huggingface [37] space. Locally, we developed a tracking system using the Unity 3D Engine, complemented by a Flask server to handle communication between the local setup and the cloud infrastructure. The development of our system was structured into three parts:

**1. Image Generation System**: This component is designed to generate a 2D image utilizing the depth information captured from the real world along with the user's text prompt.

**2. Transformation System:** This component focuses on converting the primary object from the generated image into a rational 3D model.

**3. Real-Time 3D Model Anchoring:** The final stage involves anchoring the generated 3D model onto the real object in real time.

### 3.1 Image Generation

Stable Diffusion (SD) was used as the principal tool for this task due to its robust performance, the flexibility of various checkpoint options, and its powerful integration with ControlNet. For our implementation, we deployed Stable Diffusion V2.0 with ControlNet V1.1 on our cloud server, communicating with the local setup via designated endpoints.

Within our Unity-based development environment, I added a streamlined user interface that features a simple input area. After users complete their input, the UI guides them to focus on the physical object they intend to transform. This action simultaneously triggers the depth camera to capture a depth map of the selected object. The color depth map captured by the depth camera needs to be converted to grayscale to be processed by ControlNet. Subsequently, this post-processed depth map, along with the user's textual input, forms the input for the Stable Diffusion process.

An alternative way to add depth control via ControlNet is capturing an RGB picture instead of a depth map and utilizing ControlNet's native depth estimation [38]. This eliminates the need for grayscale conversion but introduces a delay for depth inference. Some of the outcomes from this method are illustrated in Figure 5. In our comparative tests, both the traditional and alternative methods required approximately the same amount of time, but the depth estimation from RGB images could yield less accurate results, especially with unusual object shapes or under poor lighting conditions.

Multiple control types are offered in ControlNet (figure 3). Our analysis indicates that the depth method most accurately captures an object's spatial structure. This conclusion become self-evident when compared to alternative approaches, such as the segmentation method. While segmentation can accurately trace the 2D outline of an object, it often disrupt the spatial relationships. Such discrepancies often lead to inaccuracies in the subsequent CAD models according to the real object, affecting the integrity of the virtual representation.

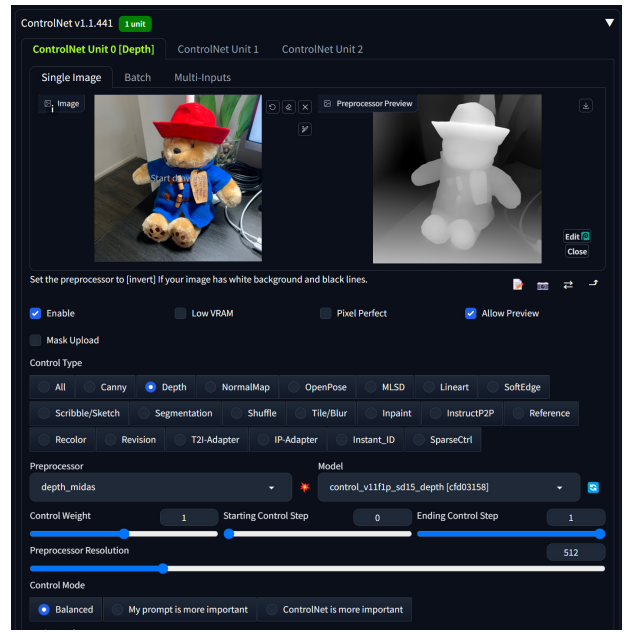

**Figure 3 The WebUI Control Panel for the Stable Diffusion ControlNet, featuring control type, control mode and more.**

ControlNet additionally features a control mode option (figure 3) that allows prioritization of either ControlNet's parameters or the user's prompt during generation. Our tests have shown that selecting the balanced mode provides the most convincing results. Prioritizing the prompt over ControlNet can lead to haptic-visual mismatches, whereas favoring ControlNet might degrade the quality of the generated images and lead to less rational outcomes.

A base model version 1.5 [39] checkpoint served as the default for SD due to its performance across a diverse array of tasks. We also evaluated customized models [37], [40] specializing in architecture and cartoon-styled figure generation and more. The results indicate that checkpoint selection can be tailored to specific application scenarios. Furthermore, common techniques in SD such as Lora[41] and Variational Autoencoders (VAE) could be applied as long as they do not compromise the spatial accuracy ensured by ControlNet.

### 3.2 3D Transformation

This part aims to construct a 3D model based on the image generated in the previous step. The processes of Image generation and 3D transformation occur sequentially in a cloud coroutine, preventing any heavy computational load from affecting the main thread in the headset.

To clarify, two distinct approaches can be utilized for achieving text-guided 3D generation with spatial control. The first approach, as detailed in this paper, involves pipelining a Text-to-Image model with an Image-to-3D model. The second approach employs direct Text-to-3D models to generate 3D meshes or point clouds[42], [43], [44] While this method reduces the data transfer between models, it lacks strong shape-control techniques, which is a crucial limitation for our applications. Although state-of-the-art techniques have shown promising capabilities for shape control with certain objects like chairs and desks [43], describing a random object's shape using only text remains challenging and counterintuitive in our use case. In contrast, ControlNet has proven to be an invaluable tool within our system. Experiments indicate that the dual-model pipeline, which combines Text-to-Image and Image-to-3D models, delivers significantly improved results in terms of spatial shape control.

TripoSR [32] was selected as our approach to realize this 3D transformation after evaluating a series of open-source models[31], [33]. We chose TripoSR due to its exceptional performance in inferring the 3D structure, its balance of speed and quality, and its native support for exporting in GLB and OBJ formats, providing the necessary performance and flexibility required for effective 3D modelling in our projects.

Before initiating the mesh generation process, the background of the image is removed using the Rembg tool [45]. This crucial step ensures that only the primary object in the scene is processed to 3D. The successful removal of the background is vital as any residual elements can shift the center of the generated model, potentially causing errors in the tracking system and leading to the mismatch between the real and virtual objects, as shown in Figure 4. For our test setups, we maintain clean or simple backgrounds to facilitate easier segmentation. Looking ahead, the integration of advanced segmentation tools [46] could enhance user interaction with the system by allowing them to selectively transform specific parts of the scene, thus improving the overall user experience.

### 3.3 Object Tracking

The third part of the system involves utilizing a model-based tracking method to accurately track the object and anchor the generated virtual model onto it in real time. This crucial step is essential for ensuring that the 3D model precisely overlays the real object, thereby effectively facilitating the visual transformation. This alignment is vital for maintaining the illusion of reality and enhancing the immersive experience, allowing users to interact with both real and virtual elements seamlessly.

For this stage, the commercial solution Vuforia [35] was deployed in Unity due to its reliable tracking capabilities and seamless integration with the game engine. The drawback is the limited customization space for the non-distribution version. While open-source solutions [34], [47] in this field offer compelling results, they also present challenges in deployment and integration, making them more suitable for skilled engineers or research teams who can navigate and manage these intricacies.

After receiving the CAD model from the cloud server, the local environment employed Vuforia's Model Target Generator software to create a tracking reference file. Integrating the target generator API could further automate this process. As illustrated in Figure 4, the image of the original object—not the image generated by Stable Diffusion—is used to generate the tracking reference. This approach also capitalizes on the original object's color information to enhance tracking accuracy, ensuring that even if the quality of the generated 3D model is suboptimal, the integrity of the tracking remains unaffected.

Once the tracking reference file is imported, it collaborates with the Vuforia plugin to accurately track the real object in 6-DoF. We then attach the generated 3D model to the runtime mesh and align their location parameters to achieve precise matching. As a result, the physical object is seamlessly rendered as its virtual replacement within the MR environment, serving as a haptic prop. This integration ensures a cohesive and immersive experience where the virtual and physical elements are perfectly synchronized.

## 4 Validation

To validate our system's usability, we conducted preliminary tests and user experiments focusing on two key perspectives:

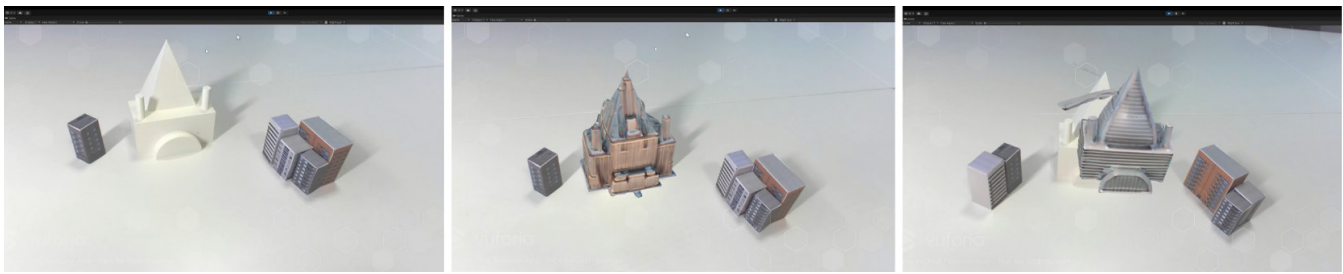

**Figure 4 Comparison of Tracking Results. Left: Original objects; Middle: Functional tracking approach; Right: Shift in tracking due to incomplete background removal.**

**1. Dynamic Transformative Capability:** Assessing whether the system can convincingly transform the physical object into any user-desired virtual forms, preserving the original 3D structure and identifying the influential factors inherent in this process.

**2. User Engagement and Interaction**: Assessing how the system enhances the user's active participation and their direct interactions with the virtual assets. To determine whether the system actively encourages users to interact more deeply with the MR environment through tangible, responsive feedback.

### 4.1 Preliminary Test

Before initiating real-user experiments, preliminary insights were gathered through a series of 2D image generation tests using Stable Diffusion and ControlNet, as depicted in Figure 5. We selected eight everyday physical objects and photographed each to create a variety of imaginative virtual assets. This process allowed us to preliminarily assess the system's transformative capability. For each prompt, we experimented with different seed numbers and chose the best result based on reasonableness and shape rationality from the first three attempts. The default v1.5 checkpoint and ControlNet v1.1 depth control were applied in this test. A balanced mode that equally prioritizes the input prompt and ControlNet's parameters were checked.

The first conclusion from our study is that transforming an object with a shape similar to the prompts generally yields more reasonable outcomes. For example, converting a basketball into an apple, or a bear toy into a rabbit toy, tends to be successful because the shapes are roughly analogous. However, objects with distinctive shapes, like a flower, pose a challenge for pre-trained diffusion models which struggle to depart significantly from the original form. Typically, these models blend elements of the original object with features of the input prompt. For instance, when a typical bottle image is used with the prompt "an apple," the model is more likely to produce an image of a bottle adorned with apple textures rather than transforming it into a standalone apple shaped like a bottle. If the "my prompt is more important" option is enabled, the model might produce something closer to a natural apple, but at the cost of losing the original object's shape control.

The second conclusion is that prompts inherently contain varied and complex shape options that can more easily lead to a plausible transformation. Prompts that invoke complex shapes like "a spaceship" or "a robot" generate a wider array of shapes compared to simpler objects like "an apple." Although there is currently no quantitative method to measure the complexity that different prompts introduce to diffusion models, it is clear that prompts involving intricate shapes tend to produce a broader spectrum of forms. However, our experiments also indicate that the richer the variety of shapes a prompt suggests, the more likely it is to generate unreasonable details. This finding opens an intriguing pathway for further exploration in prompt engineering within this workflow, potentially leading to enhanced capabilities and more controlled outcomes in image generation through refined prompt structuring.

### 4.2 Usability Study

#### 4.2.1 Test settings

To comprehensively test the system's usability, a structured usability study was designed to evaluate the capacity of participants to convert a physical object into a corresponding virtual representation. (Figure 6) The study recruited nine individuals—five males and four females—all of whom had previous experience with VR devices. The experiment was set up in a clean seminar room to eliminate external distractions. The Oculus Quest 2 headset paired with a ZED Mini depth camera system was used in the test, with a typical Paddington Bear toy serving as the haptic prop.

The test began with a brief about the project's objectives without introducing specific examples to reduce preconceived notions or bias. Participants were then equipped with the headset and given two minutes to get used to the VR environment, user interface, input system and the bear toy.

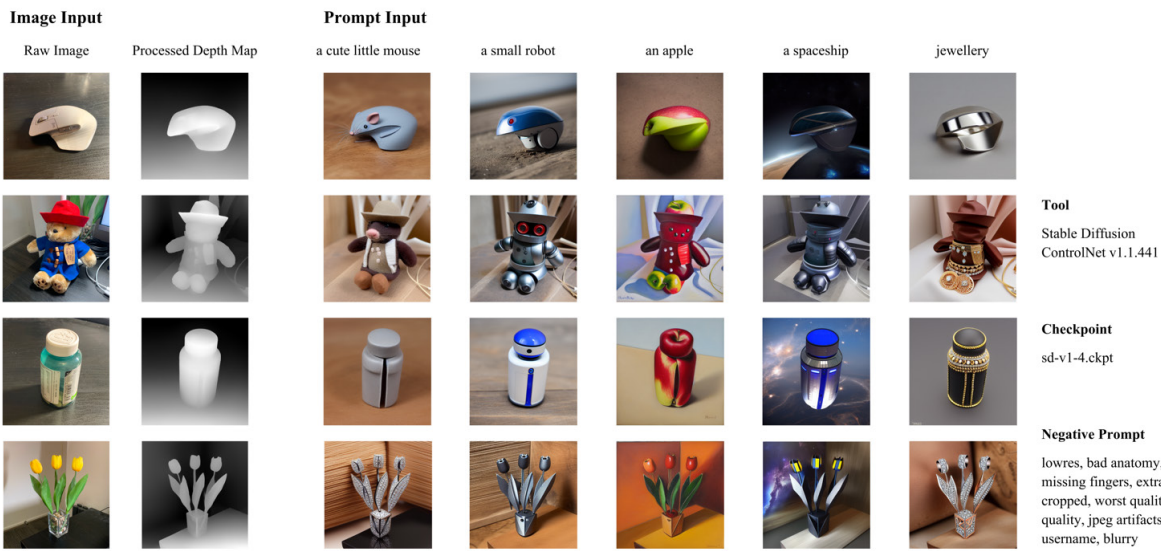

**Figure 5 Partial Results from Preliminary Tests Using Stable Diffusion and ControlNet.**

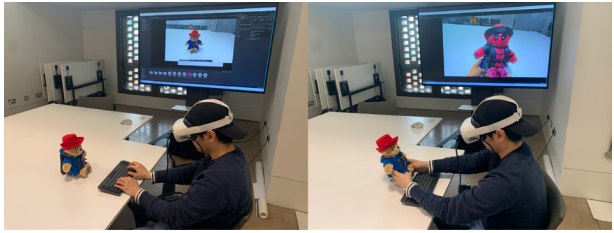

**Figure 6 Usability Study.** Left: Participant enters his prompt 'Deadpool' via keyboard; Right: Participant interacts with the generated result.

Following this initial familiarization, participants were instructed to start the process by typing their prompts in the headset. Each participant was given three times to try different prompts while their inputs and reactions were recorded (Figure 7). After completing the test and removing the headset, participants completed a structured survey comprising of five questions. Four of these questions required responses on a seven-point scale, with one open-ended question.

The specific statements in the survey were as follows:

Q1: "The transformation result meets my expectations (overall and for each try)."

Q2: "The generated object appears realistic, as if they were truly there."

Q3: "The experience feels captivating and intriguing."

Q4: " I'm interested in applying this technology to my own items and settings."

Additionally, an open-ended question (Q5) was included: "Where else would you like to apply this technology?"

After the survey, we conducted post-questionnaire interviews to delve deeper into participants' experiences. These interviews were aimed at gathering qualitative insights into the various factors that influenced their interactions with the system. This step allowed us to explore in more detail the subjective perceptions and specific feedback from each participant.

### 4.2.2 Results and Discussion

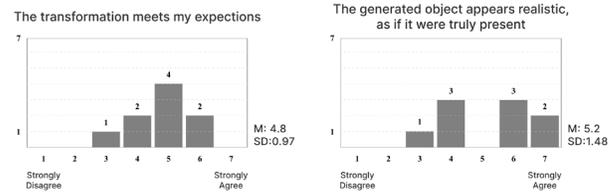

**Figure 8 Results of Question 1 and 2**

The outcomes from the initial question (Figure 8) show that the transformation result aligned with participants' expectations to a moderate degree, evidenced by an average score of 4.8 with a standard deviation of 0.97. This suggests a generally positive response, although the spread indicates some variance in satisfaction levels among participants.

Further analysis of responses to the second question returned an even higher average score of 5.2, with a larger standard deviation of 1.48, indicating a larger variability in the perceived realism of the transformation. Interviews revealed that this variability was closely linked to the quality of the generative process. Lower ratings were predominantly associated with issues such as irrational generations, which included problems like unreasonable textures, broken meshes, and poor tracking results.

Additional insights emerged during the post-survey discussions. A marked factor was the "low refresh rate" of the depth camera embedded with the headset, capped at 15 fps at its widest field of view. This limitation led to a noticeable "laggy" effect that diminished the immersiveness. Moreover, the issue of "loss of tracking" was also mentioned, this typically occurred when participants unintentionally obstructed the camera's view of the bear by holding it from the front or moving it too rapidly, resulting in the system losing track of the object and causing the virtual model to freeze momentarily. This issue was partially mitigated after instructions on how to hold the bear to avoid blocking it were given.

The analysis of generative results illustrates a clear pattern in how the nature of prompts influenced the perceived success of transformations. Prompts that closely matched the original

| | | |
|---|---|---|
| Casper (2) (B) | a Casper ghost in smiling face (3) (B) | Teddybear holding a rocket and spear weapon (4) (A) |
| Spaceship (3) (C) | silver spaceship with sharp head and blue tail (4) (C) | Eiffel Tower (2) (B) |
| Green Goblin from the original Spiderman movie (4) (A) | Italian woman screaming bloody murder (5) (A) | Tin Foil (6) (C) |
| sport car (3) (B) | a cute shark child toy (5) (B) | a cute shark doctor child toy (5) (B) |
| magician (3) (A) | a witch wearing a hat (4) (A) | a pink sheep (7) (B) |
| a blue Smurfs wearing hats (5) (A) | a boy wearing a suit (6) (A) | Appleman (5) (C) |
| a cowboy barbie in all pink (5) (A) | an alien from Jupyter with four eyes (7) (C) | a VC jelly gummy in orange flavour (6) (C) |
| astronaut (7) (A) | Alien bear (4) (A) | Peppa pig (5) (A) |
| deadpool (7) (A) | SpongeBob (6) (B) | Certain Celebrity*(name hidden) (5) (A) |

**Figure 7 Recorded prompts from Our Usability Test.** Each row represents three attempts by each participant, where the first bracketed content indicates the rating received, and the second bracketed content specifies the group classification based on the nature of the prompt.

object's shape or were culturally iconic (e.g., 'Green Goblin', 'SpongeBob', 'Deadpool') were associated with higher satisfaction ratings. Conversely, prompts that greatly diverged from the physical prop's form, such as 'Eiffel Tower', typically resulted in lower scores and less rational results. To systematically evaluate, prompts were categorized into three distinct groups based on their resemblance to the haptic prop:

*Group A*: Prompts closely resembling the prop in both shape and size.

*Group B*: Prompts with no resemblance to the prop in shape or size.

*Group C*: Prompts encompassing a broad spectrum of shapes and sizes.

The analysis yielded the following results:

*Group A*: Average score of 4.9 with a standard deviation of 1.12, indicating relatively consistent satisfaction.

*Group B*: Lower average score of 4.1, accompanied by a higher standard deviation of 1.76, reflecting significant variability and generally poorer outcomes.

*Group C*: A high average score of 5.2 with a standard deviation of 1.28, suggesting that diverse and imaginative transformations were well-received.

These statistics highlight the importance of prompt relevance to the physical characteristics of the haptic prop in achieving satisfactory transformations. The highest ratings in Group C also suggest that participants may have approached the task with a sense of exploration and openness to varied outcomes as many participants said these results were "not exactly what I expected but was quite interesting." The results of the experiment demonstrate that imaginative prompts can lead to surprisingly high-quality and innovative results, exemplified by prompts such as 'Appleman' (rated 5) and 'Tin Foil' (rated 6).

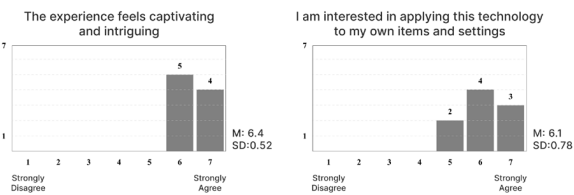

**Figure 9 Results of Question 3 and 4**

The participant feedback captured through the usability study indicates a positive reception of the transformation experience, which was consistently described as "interesting" and "captivating." Question 3 received a mean score of 6.4 and a standard deviation of 0.52. Participants also expressed interest in the potential applications of this technology to their personal items and environments, with a mean score of 6.1 and a standard deviation of 0.78 for question 4 (Figure 9).

Post-survey discussions revealed a broad spectrum of personal and creative applications envisioned by participants. Over half of them promptly identified practical uses in their daily lives, ranging from converting everyday objects like "pillows" and "bikes" into more aesthetic or functional forms, to transforming personal spaces such as "my apartment to a luxury villa". Additionally, more imaginative applications were suggested, including the transformation of personal items into representations of celebrities or fictional characters, such as transforming a bear toy into "Taylor Swift" or "James Bond." Others envisioned specific customizations, such as altering their "Bearbrick toys" [48] into various styles or leveraging the technology for rapid design prototyping. These responses not only illustrate the versatility of the technology but also highlight its potential to profoundly enhance both personal expression and professional creative workflow.

In conclusion, the usability study provided validation of our system's transformation capability to seamlessly convert the physical object into any user-desired virtual forms. Several key factors influencing user experience including technical limitations were pointed out to help further investigation. Participants also demonstrated considerable enthusiasm for integrating this technology into their own creative processes and everyday environments. This strong interest underscores the system's potential scenarios and its capability to inspire innovation across various fields. The study also suggested important correlations between the nature of the prompts and the original physical objects

## 5 Limitations

**Object Tracking Solutions.** The method for object tracking has relied on commercial solutions, which introduce watermarks and impose constraints on customization and automation. More advanced tools could be applied with more skilled engineers or research teams. Additionally, the current implementation of the transformation process does not fully occur within a true 3D space. The tracking component primarily processes 2D video data and simply overlays the virtual 3D model onto the detected surface within the video frame. This technique has considerable shortcomings, particularly the absence of realistic occlusion effects. This limitation critically undermines the authenticity and immersive quality of the transformation, making the interaction between real and virtual objects appear less convincing.

Improvements in these areas are crucial for advancing the system's ability to provide more realistic and interactive virtual experiences. By transitioning to a more sophisticated tracking system that operates in genuine 3D space and addresses the current occlusion limitations, the technology could achieve a greater degree of realism and utility, thereby enhancing user engagement and satisfaction.

**Immersiveness.** Our setup does not support a 360-degree stereo pass-through augmented reality (AR) effect. Instead, a depth camera is mounted in front of the headset, aligned with the direction the user's head is facing. The captured video, after undergoing processing, is streamed into the headset, enabling users to view their surroundings. However, this setup does not replicate a stereo AR pass-through effectively. The resultant user experience is viewing a large cinema-like screen within the headset that streams real-time post-processed video, as shown in Figure 10. Privacy concerns restrict most headset manufacturers from providing full access to camera and depth sensor capabilities to Indi-developers. Efforts have been made to mitigate these limitations by optimizing the alignment of the screen's position relative to the user's eye. Nonetheless, achieving a more immersive and interactive experience would greatly benefit from access to more open hardware platforms or greater engineering resources, allowing for the development of a more sophisticated and capable AR system.

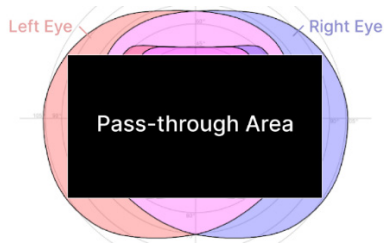

**Figure 10 The Display area shown in the headset.**

**Usability test.** The usability tests conducted as part of this research included a small sample size of only nine participants, which may not provide a comprehensive representation of diverse user experiences. Additionally, the methods employed in both the preliminary tests and the usability studies involved subjective elements in prompt data processing and interpretation, introducing potential biases. In future research, more objective, quantifiable metrics for evaluating system performance and user engagement could be involved.

## 6  Future Work

**User Experience.** To improve user experience, several enhancements can be made to the system's interface and interaction methods. A clearer and more desirable interface could be designed to provide clearer guidance on how to interact with the system, such as where to position objects and remain still during capture. Incorporating voice commands could replace peripheral input devices, making the system more intuitive. Additionally, displaying progress indicators for each transformation step and integrating advanced segmentation tools would help users more effectively manage complex backgrounds and select the primary object for transformation.

**Creativity and Imagination.** Our experiments and surveys highlight a promising potential for leveraging this technology in Artificial Intelligence for General Creativity (AIGC). Usability studies also demonstrated a strong interest from participants in incorporating the system into their home settings or workflows during the survey and interview sessions. Additionally, the creative prompts in the test underscore the technology's capacity to inspire and facilitate creativity across various domains. New possibilities and understandings could be unlocked by further exploring diverse application scenarios.

**Model Coupling.** The current approach to pipeline the models includes sequential use of available open-source models, which may not optimally utilize the potential richness of input data. Specifically, our process involves using a single picture or depth map as the initial input, which feeds into a Text-to-Image model. The output from this model subsequently serves as the input for an Image-to-3D model. This linear and isolated application, while functional, leverages only a small fraction of the information that our hardware is capable of providing. In recent advancements in 3D reconstruction [49], views from multiple angles were estimated to facilitate the generation process. Providing this information directly with the depth camera could potentially improve the accuracy and quality of the final model. Incorporating a more interconnected model coupling strategy to enable more input information could substantially enhance the accuracy and quality of the outputs.

**AI-Guided Transformation.** Our system currently necessitates that users specify their transformation goals explicitly. However, integrating Large Language Models like ChatGPT could streamline and enrich this process. For instance, as demonstrated in Figure XX, ChatGPT was used to describe the tactile sensation of a pillow in just four words. Such descriptive outputs could be incorporated as inputs into our generative AI system, enabling it to produce more accurate and lifelike transformations that reflect the true characteristics of the real objects. This approach not only simplifies user interaction but also enhances the creative potential of the system by leveraging sophisticated natural language processing to interpret and materialize user intentions.

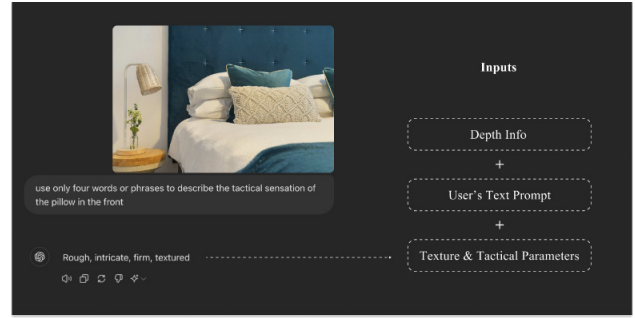

**Figure 11 Conceptualization of enhancing generation immersiveness using large language models by incorporating estimated tactile sensations of real objects as part of the input.**

**Multiple Objects & Spatial Transformation.** In the current setup, transformations are limited to individual objects due to tracking constraints. However, there is potential to expand this capability to handle multiple objects simultaneously, enabling spatial-level transformations. This would involve sequentially transforming different objects within a space while maintaining continuous tracking of each item.

Imagine a scenario where a user could interact with a virtual assistant within their room. By simply instructing, "Transform this room into a space cabin style," the system could autonomously adjust each object in the room to match the desired theme, adapting not only to the overall style but also to the unique characteristics of each item. Implementing this capability would require advanced techniques to accurately identify and process multiple items. This would also represent a significant leap forward and offer new possibilities for personal entertainment, creative workflow and more.

## 7  Conclusion

Haptic Repurposing with GenAI introduces a novel approach to transforming the everyday tangible world into dynamic, interactive haptic interfaces for virtual assets using generative AI. Throughout this paper, we have detailed our development process, acknowledged limitations, and highlighted areas for future iterations. Our usability tests have evaluated the system's effective transformation capabilities and its ability to enhance user engagement and authenticity in Mixed Reality experiences. Additionally, these tests have demonstrated a clear potential for fostering creativity. The hope is that this project will pave the way for more natural and intuitive interactions within digital environments, driving further innovation and the broader integration of MR technologies into everyday life.